\documentclass[english,10pt,journal,comsoc]{IEEEtran}
\usepackage[T1]{fontenc}
\usepackage[latin9]{inputenc}
\usepackage{color}
\usepackage{array}
\usepackage{multirow}
\usepackage{amsmath}
\usepackage{graphicx}

\makeatletter

\providecommand{\tabularnewline}{\\}

\usepackage{booktabs}
\usepackage{float}

\usepackage{rotating}
\usepackage{array}
\newcolumntype{L}[1]{>{\vspace{0.5em}\begin{minipage}{#1}\raggedright\let\newline\\
\arraybackslash\hspace{0pt}}m{#1}<{\end{minipage}\vspace{0.5em}}}
\newcolumntype{R}[1]{>{\vspace{0.5em}\begin{minipage}{#1}\raggedleft\let\newline\\
\arraybackslash\hspace{0pt}}m{#1}<{\end{minipage}\vspace{0.5em}}}
\newcolumntype{C}[1]{>{\vspace{0.5em}\begin{minipage}{#1}\centering\let\newline\\
\arraybackslash\hspace{0pt}}m{#1}<{\end{minipage}\vspace{0.5em}}}

\makeatother

\usepackage{babel}
\begin{document}
\title{Learning to Score Figure Skating Sport Videos}

\author{ Chengming~Xu,   Yanwei~Fu,  Zitian Chen,Bing Zhang, Yu-Gang Jiang, Xiangyang Xue  
\IEEEcompsocitemizethanks{ 
\IEEEcompsocthanksitem Chengming~Xu , Bing Zhang,and Yanwei~Fu are with the School of Data Science, Shanghai Key Lab of Intelligent Information Processing, Fudan University, Shanghai, China. Email: \{18110980002, yanweifu\}@fudan.edu.cn.  
\IEEEcompsocthanksitem Yanwei Fu is the corresponding author.
\IEEEcompsocthanksitem  Zitian Chen, Yu-Gang Jiang and Xiangyang Xue are with the School of Computer Science, Shanghai Key Lab of Intelligent Information Processing, Fudan University. Email: \{ygj,xyxue\}@fudan.edu.cn.   }
\thanks{}}  

\maketitle

\begin{abstract} 

This paper aims at learning to score the figure skating sports videos. 
To address this task, we propose a deep architecture that includes two complementary components, i.e.,  Self-Attentive LSTM and Multi-scale Convolutional Skip LSTM. These two components can efficiently learn the local and global sequential information in each video. Furthermore,  we present a large-scale figure skating sports video dataset -- FisV dataset. This dataset includes
500 figure skating videos with the average length of 2 minutes and 50 seconds. Each video is annotated by two scores of nine different referees, i.e., Total Element Score(TES) and Total Program Component Score (PCS). 
Our proposed model is validated on FisV and MIT-skate datasets. The experimental results show the effectiveness of our models in learning to score the figure skating videos.

\end{abstract}


\section{Introduction}

With the rapid development of digital cameras and proliferation of
social media sharing, there is also an explosive growth of available
figure skating sports videos in both the quantity and granularity.
Every year there are over 20 international figure skating competitions
held by International Skating Union (ISU) and hundreds of skaters
participated in them. Most of the high-level international competitions,
such as ISU championships and ISU Grand Prix of Figure Skating are
broadcast on the worldwide broadcaster, for instance CBC, NHK, Eurosport,
CCTV. Over 100 figure skating videos are uploaded in Youtube and Dailymotion
a day during the season. 

The analysis of figure skating sports videos also have many real-world
applications, such as automatically scoring the players, highlighting
shot generation, and video summarization. By the virtue of the state-of-the-art
deep architectures and action recognition approaches, the techniques
of analyzing figure skating sports videos will also facilitate statistically
comparing the players and teams, analyzing player's fitness, weaknesses
and strengths assessment. In terms of these sport statistics, professional
advice can be drawn and thus help the training of players. 

Sports video analytics and action recognition in general have been
extensively studied in previous works. There exist many video datasets,
such as Sports-1M \cite{deepvideo}, UCF 101\cite{Soomro2012}, HMDB51
\cite{HMDB51}, FCVID \cite{fcvid_2017} and ActivityNet \cite{activitynet}.
These datasets crawled the videos from the search engines (\emph{e.g.},
Google, or Bing) or the social media platforms (e.g., YouTube, Flickr,
\emph{etc}). The videos are crowdsourcingly annotated. On these video
datasets, the most common efforts are mainly made on video classification
\cite{jiang2011consumervideo,deepvideo}, video event detection \cite{over2011trecvid},
action detection and so on. 

Remarkably, inspired by Pirsiavash \emph{et al}. \cite{quality_action},
this paper addresses a novel task of learning to score figure skating
sport videos, which is very different from previous action recognition
task. Specifically, in this task, the model must understand every
clip of figure skating video (\emph{e.g.}, averagely 4400 frames in
our Fis-V dataset) to predict the scores. In contrast, one can easily
judge the action label from the parts of videos in action recognition.
For example, one small video clip of capturing the mistake action
of the player, will significantly negatively affect the final scores
in our task. Thus the model must fully understand the whole video
frames and process the varying length of videos.

Quite a few works have been devoted to learning to score figure skating
videos.  The key challenges come from several aspects. First, different
from consumer videos, figure skating videos are the professional sports
videos with the longer length (averagely 2 minutes and 50 seconds).
Second, the scores of figure skating videos should be contributed
by the experts or referees; in contrast, the labels of previous classification/detection
based video analysis tasks are collected in a crowdsourcing way. Third,
not all video segments can be useful to regress the scores, since
the referees only take account into scores those clips of technical
movements (TES) or a good interpretation of music (PCS). 

\begin{figure*}
\begin{centering}
\includegraphics[scale=0.3]{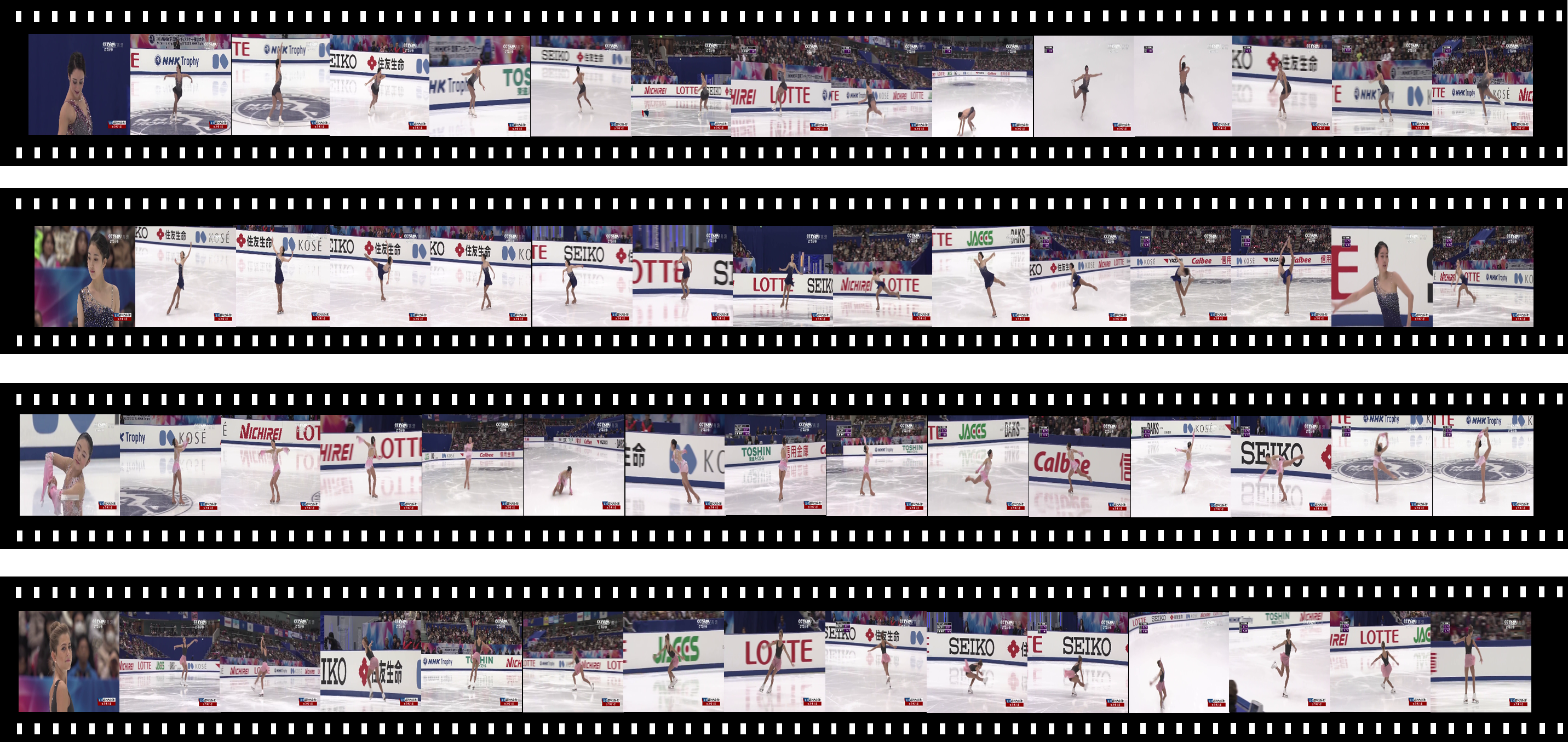}
\par\end{centering}
\caption{\emph{\label{fig:Example-frames-of}}Example frames of our figure
skating video dataset. Each row is corresponding to one video.}
\end{figure*}

To address these challenges, we propose an end-to-end framework to
efficiently learn to predict the scores of figure skating videos.
In particular, our models can be divided into two complementary subnetworks,
\emph{i.e.}, Self-Attentive LSTM (S-LSTM) and Multi-scale Convolutional
Skip LSTM (M-LSTM). The S-LSTM employs a simple self-attentive strategy
to select important clip features which are directly used for regression
tasks. Thus the S-LSTM mainly learns to represent the local information.
On the other hand, the M-LSTM models the local and global sequential
information at multi-scale. In M-LSTM, we utilize the skip LSTM to
efficiently save the total computational cost. Both two subnetworks
can be directly used as the models for prediction, or integrated into
a single framework for the final regression tasks. Our models are
evaluated on two figure skating datasets, \emph{namely}, MIT-skate
\cite{quality_action} and our own Fis-V video dataset. The experiments
results validate the effectiveness of our models.

To further facilitate the research of learning to score figure skating
videos, we contribute the Figure Skating Video (Fis-V) dataset to
the community. The Fis-V dataset has the videos of high quality as
well as the scores labeled. Specifically, our videos in Fis-V are
captured by professional camera devices. The high standard international
figure skating competition videos are employed as the data source
to construct Fis-V dataset. The example video frames of this dataset
are shown in Fig. \ref{fig:Example-frames-of}. Each video snapshots
the whole performance of one skater only; the irrelevant parts towards
the skater (such as warming up, bowing to the audience after the performance)
are pruned. Thus the length of each video is about 2 minutes and 50
seconds. Totally, we collect 500 videos of 149 professional figure
skating players from more than 20 different countries. We also gather
the scores given by nine different international referees in the competitions.

\vspace{0.1in}

\noindent \textbf{Contributions. }We highlight the three contributions.
(1) The proposed Self-Attentive LSTM can efficiently learn to model
the local sequential information by a self-attentive strategy. (2)
We propose a Multi-scale Convolutional Skip LSTM model in learning
the local and global information at multi-scale, while it can save
the computational cost by skipping some video features. (3) We contribute
a high quality figure skating video dataset \textendash{} Fis-V dataset.
This dataset is more than 3 times bigger than the existing MIT-skate
dataset. We hope this dataset can boost the research of learning to
score professional sports videos. 

The rest of this paper is organized in such a way. Sec. \ref{sec:Related-Work}
compares some related work. We describe the details of constructing
the dataset in Sec. \ref{sec:Figure-skating-Video}. The methodology
of solving the specific task is discussed in Sec. \ref{sec:Methodology}.
We finally give the experimental results in Sec. \ref{sec:Experiments}.
The whole paper is concluded in Sec. \ref{sec:Conclusion}.  

\section{Related Work\label{sec:Related-Work}}

The sheer volume of video data makes the automatic video content understanding
difficulty intrinsically. Very recent, deep architectures have been
utilized to extract feature representations effectively in the video
domain. While the development of image representation techniques
has matured quickly in recent years \cite{gupta_pami2009,graph_matching,human_still_latent,distributed_repre_cvpr2011,person_interaction},
more advanced architectures were conducted for video understanding
\cite{retrieving_movie,action_bank,efros2003action}, including Convolutional
Networks (ConvNets) with Long Short-Term Memory (LSTMs) \cite{donahue_cvpr2015,snippet_cvpr2015}
and 3D Convolutional Networks \cite{two_strems_nips} for visual recognition
and action classification, two-stream network fusion for video action
recognition \cite{snippet_cvpr2015,two_strems_nips}, Convolutional
Networks learning spatiotemporal features \cite{taylor_eccv2010,Tran_ICCV2015}.
We discuss these previous works in each subsection. 

\subsection{Video Representation}

Previous research on improving video representations focuses on local
motion features such as HOF \cite{laptev2008actions}{[}24{]} and
MBH \cite{MBH} in the Dense Trajectories feature \cite{dense_traj}
and the corresponding variants \cite{improved_dense}. The success
of deep learning in video analysis tasks stems from its ability to
derive discriminative spatial-temporal feature representations directly
from raw data tailored for a specific task ~\cite{Ji2010,C3D}.  

Directly extending the 2D image-based filters \cite{KrizhevskySH12}
to 3D spatial-temporal convolutions may be problematic. Such spatial-temporal
Convolutional Neural Networks (CNN), if not learned on large-scale
training data, can not beat the hand-crafted features. Wang \emph{et
al}. \cite{Wang2013a} showed that the performance of 3D convolutions
is worse than that of state-of-the-art hand-crafted features. Even
worse, 3D convolutions are also computationally expensive; and it
normally requires more iterations to train the deep architectures
with 3D convolutions than those without. 

To reduce such computational burden, Sun \emph{et al}. proposed to
factorize spatial-temporal convolutions ~\cite{sun2015human}. It
is worth noting that videos can be naturally considered as an ensemble
of spatial and temporal components. Motivated by this observation,
Simonyan and Zisserman introduced a two-stream framework, which learn
the spatial and temporal feature representations concurrently with
two convolutional networks~\cite{Simonyan2014}.  Such a two stream
approach achieved the state-of-the-art performance on many benchmarks.
Furthermore, several important variants of fusing two streams are
proposed, such as \cite{cvpr15:wang,Feichtenhofer16,WangXWQLTV16,ZhangWWQW16,Wang_Transformation,zhu2016key,bilen2016dynamic,icmr15:eval2stream}

Most recently, C3D \cite{C3D}, and SENet \cite{SEnet} have been
proposed for powerful classification models on videos and images.
C3D \cite{C3D} utilized the $3\times3\times3$ spatial-temporal convolution
kernels and stacked them into a deep network to achieve a compact
representation of videos. C3D has been taken as a more effective structure
of preserving temporal information than 2D CNNs. SENet \cite{SEnet}
adopted the ``Squeeze-and-Excitation'' block, which integrates the
channel-set features, stressing the independencies between channels.
In this work, we employ the C3D as the basic video feature representation.

\subsection{Video Fusion}

In video categorization systems, two types of feature fusion strategies
are widely used, \emph{i.e.}, the early fusion and the late fusion.
Multiple kernel learning \cite{mkl_2004} was utilized to estimate
fusion weights \cite{heterog_iccv_2009,natarajan_2012}, which are
needed in both early fusion and late fusion. To efficiently exploit
the relationships of features, several more advanced feature fusion
techniques were conducted. An optimization framework in \cite{late_fusion_2012}
applied a shared low-rank matrix to reduce noises in the fusion. An
audio-visual joint codebook proposed by Jiang\emph{ el al.} \cite{jiang_acmmm2009}
discovered and fused the correlations of audio and visual features
for video classification. The dynamic fusion is utilized in \cite{liu_2013_cvpr}
as the best feature combination strategy. 

With the rapid growth of deep neural networks, the combination of
multiple futures in neural networks gradually comes into sight. In
multimodal deep learning, a deep de-noised auto-encoder \cite{multimodal_icml2011}
and Boltzmann machines \cite{Srivastava_NIPS2012} were employed to
fuse the features of different modalities. More recently, Recurrent
Neural Networks have also been utilized to fuse the video representation.
\textcolor{red}{}Wu \emph{et al}. \cite{wu2016multi} modeled videos
into three streams including frames, optical flow and audio spectrogram
and fuse classification scores adaptively from different streams with
learned weights. Ng \emph{et al}. \cite{pyarmid_video_fusion} employed
time domain convolution or LSTM to handle video structure and use
late fusion after the two-stream aggregation. Comparing with this
work, we propose a fusion network to efficiently fuse the local and
global sequential information learned by the self-attentive and M-LSTM
models.

\subsection{Sports Video Analysis}

Recently, the sports video analysis has been tropical in the research
communities \cite{camera_light}.  A common and important unit in
sports video analysis is the action, or a short sequence of actions.
There are various works that assess how well the people perform actions
in different sports, including an application of automated video assessment
demonstrated by a computer system that analyzes video recordings of
gymnasts performing the vault \cite{gordon_1995}; a probabilistic
model of a basketball team playing based on trajectories of all players
\cite{jug_2003}; the trajectory-based evaluation of multi-player
basketball activity using Bayesian network \cite{perse_2007}; and
machine learning classifier on top of a rule-based algorithm to recognize
on-ball screens \cite{McQueen2014}. 

The tasks of learning to score the sports have less been studied with
only two exceptions \cite{quality_action,parmar2017learning}. Pirsiavash
\emph{et al}. \cite{quality_action} introduced a learning-based framework
evaluating on two distinct types of actions (diving and figure skating)
by training a regression model from spatiotemporal pose features to
scores obtained from expert judges. Parmar \emph{et al.} \cite{parmar2017learning}
applied Support Vector Regression (SVR) and Long Short-Term Memory
(LSTM) on C3D features of videos to obtain scores on the same dataset.
In both \cite{quality_action,parmar2017learning}, the regression
model is learned from the features of video clips/actions to the sport
scores. Comparing with \cite{quality_action,parmar2017learning},
our model is capable of modeling the nature of figure skating. In
particular, our model learns to model both the local and global sequential
information which is essential in modeling the TES and PCS. Furthermore,
our self-attentive and M-LSTM model can alleviate the problem that
figure skating videos are too long for an ordinary LSTM to get processed.

\section{Figure Skating Video (Fis-V) Dataset\label{sec:Figure-skating-Video}}

Our figure skating video dataset is designed to study the problem
of analyzing figure skating videos, including learning to predict
scores of each player, or highlighting shots generation. This dataset
would be released to the community under necessary license.

\subsection{Dataset construction}

\begin{figure*}
\begin{centering}
\includegraphics[scale=0.4]{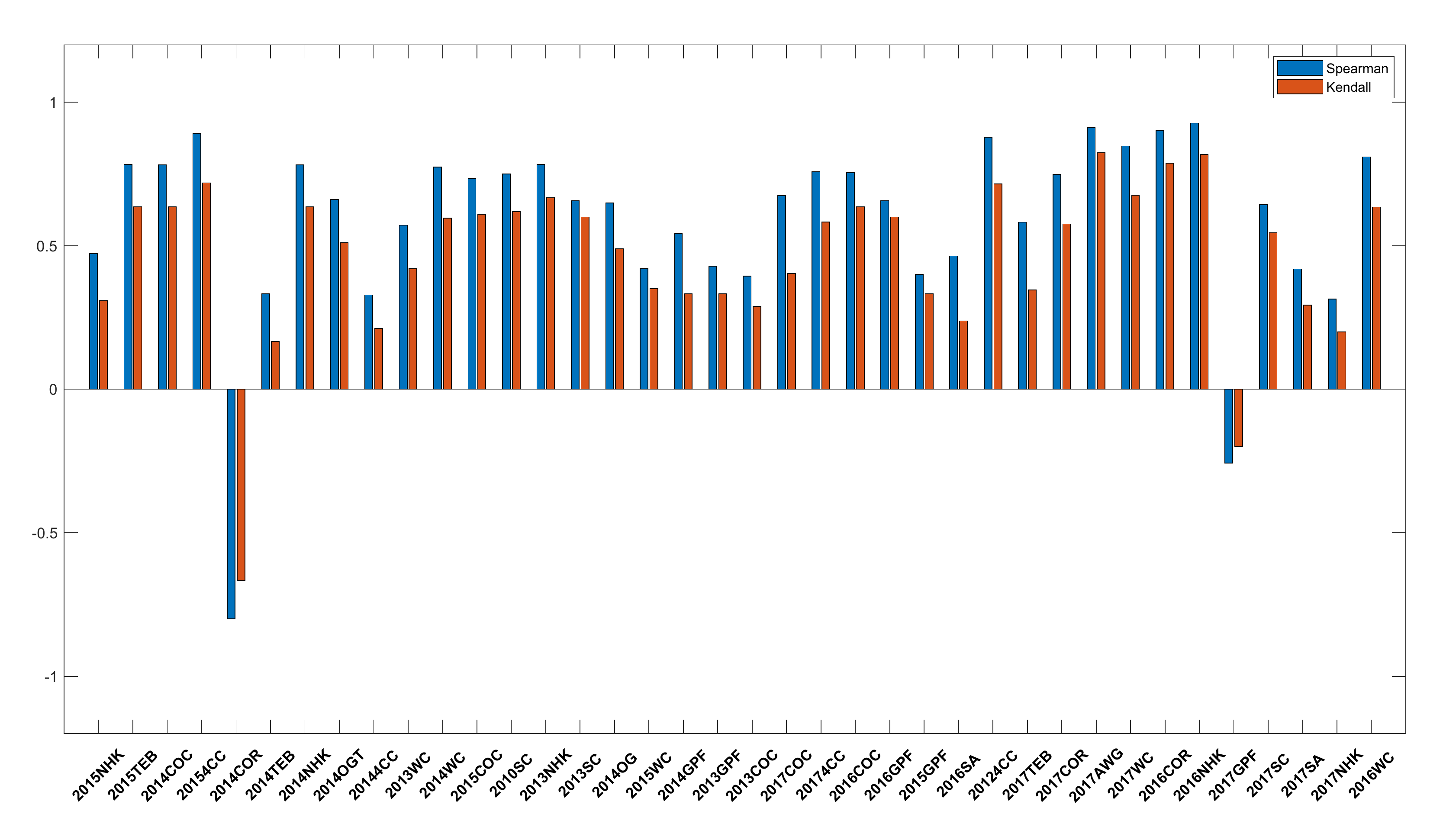}
\par\end{centering}
\caption{\label{fig:matches}Results of the correlations over different matches.
The correlation values are visualized in Y-axis; and the X-axis denotes
different matches. For instance, ``13COC'' means the COC (Cup of
China) held in 2013.}
\end{figure*}

\noindent \textbf{Data source}. To construct the dataset, we search
and download a great quantity of figure skating videos. The figure
skating videos come from formal high standard international skating
competitions, including NHK Trophy (NHK), Trophee Eric Bompard (TEB),
Cup of China (COC), Four Continents Figure Skating Championships (4CC)
and so on. The videos of our figure skating video dataset are only
about the playing process in the competitions. Note that the videos
about figure skating may also be included in some previous datasets
(\emph{e.g.}, UCF 101\cite{Soomro2012}, HMDB51 \cite{HMDB51}, Sports-1M
\cite{deepvideo} and ActivityNet \cite{activitynet}), which are
constructed by searching and downloaded from various search engines
(e.g., Google, Flickr and Bing, etc), or the social media sharing
platforms (e.g. Youtube, DailyMotion, etc.). Thus the data sources
of those datasets are different from ours. We thus emphasize the better
and more consistent visual quality of our TV videos from the high
standard international competitions than those consumer videos downloaded
from the Internet. Additionally, the consumer videos about figure
skating may also include the practice videos. 

\vspace{0.1in}

\noindent \textbf{Selection Criteria.} We carefully select the figure
skating videos used in the dataset. We assume the criterion of scores
of figure skating should be consistent for the high standard international
skating competitions. Thus to maintain standard and authorized scoring,
we select the videos only from the highest level of international
competitions with fair and reasonable judgement. In particular, we
are using the videos from ISU Championships, ISU Grand Prix of Figure
Skating and Winter Olympic Games. Totally we have the videos about
149 players from more than 20 different countries. Furthermore, in
figure skating competitions, the mark scheme is slightly changing
every season, and very different for men and women. To make the scores
more comparable, only the competition videos about ladies' singles
short program happened over the past ten years are utilized in our
figure skating video dataset. We also collect the ground-truth scores
given by nine different referees shown in each competition. 

\vspace{0.1in}

\noindent \textbf{Not rush videos}. The rush videos often refer to
those unedited videos, which normally contain redundant and repetitive
contents. The videos about figure skating in previous datasets may
include those unedited and ``rush'' parts about the players, such
as warming up, bowing to the audience after the performance, and waiting
for scores at the Kiss\&Cry. These parts may be not necessarily useful
to help judge the scores of the performance of figure skating. In
contrast, we aim at learning the model of predicting the scores purely
from the competition performance of each player, rather than from
the ``rush'' parts. Thus those unedited parts are pruned in our
videos. More interestingly and importantly, in the sports videos of
multiple players, the videos have to track, locate and transit different
players. Our figure skating video has about only one player, and the
whole video is only tracking, and locating the player over her whole
performance as shown in Fig. \ref{fig:Example-frames-of}.

\begin{figure*}
\begin{centering}
\includegraphics[scale=0.4]{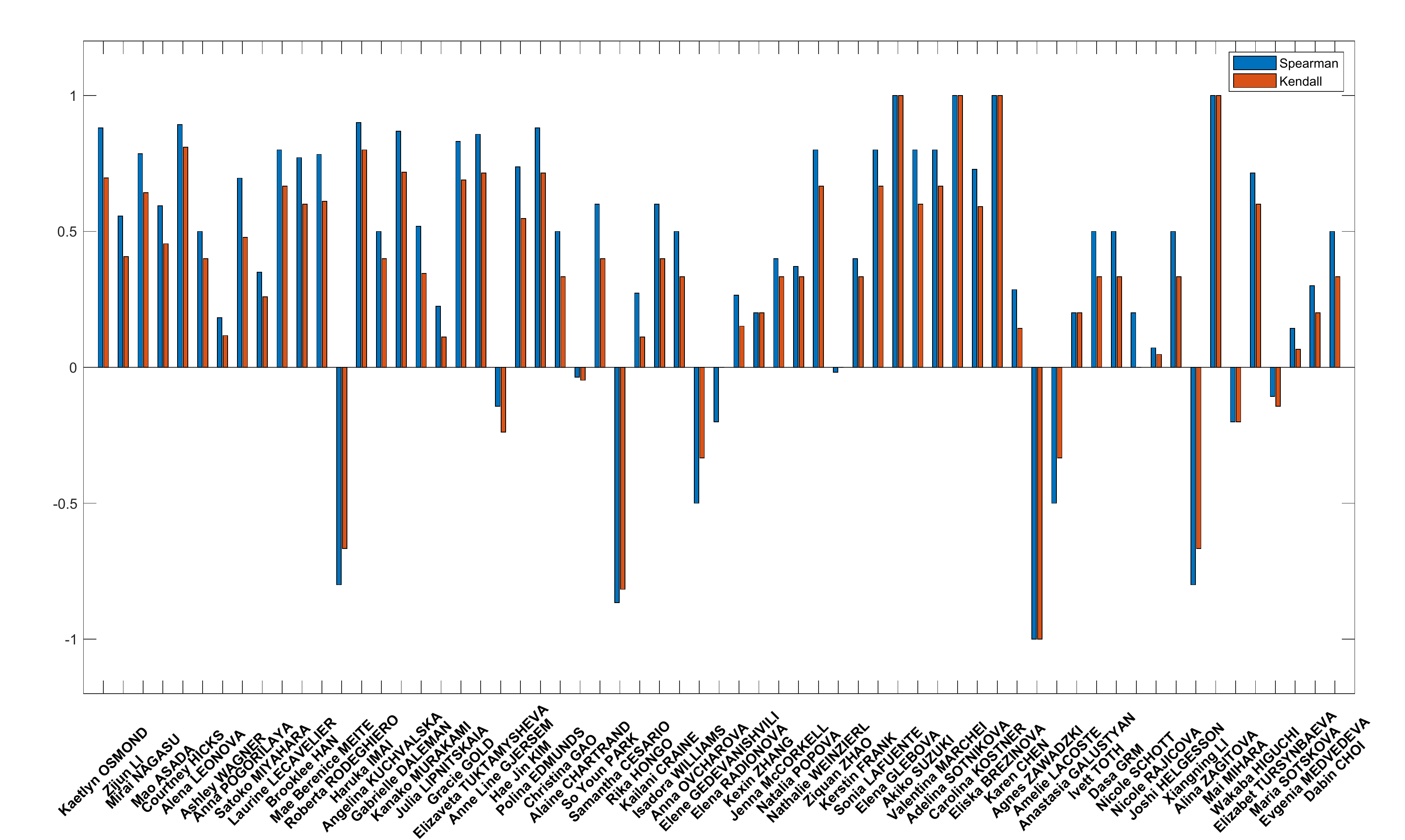}
\par\end{centering}
\caption{\label{fig:players}Results of the correlations of different players.
The Y-axis indicates the computed correlation values; and the X-axis
annotates the name of each skater. }
\end{figure*}

\subsection{Pre-processing and Scoring }

\vspace{0.1in}

\noindent \textbf{Pre-processing}. We initially downloaded 100 hour
videos; and the processing procedure is thus needed to prune some
low quality videos. In particular, we manually select and remove the
videos that are not fluent nor coherent. To make sure the figure skating
videos exactly correspond to the ground-truth scores, we manually
processed each video by further cutting the redundant clips (e.g.
replay shots or player's warming up shots). We only reserve the video
from the exact the beginning of each performance, to the moment of
ending pose, with duration of about 2 minutes and 50 seconds. Particularly,
this time slot also meets the duration of skating stipulated by the
International Skating Union, which is 2 minutes and 40 seconds within
10 seconds plus or minus for ladies\textquoteright{} singles short
program. Each video has about 4300 frames with the frame rate 25.
Thus both the number of frames and videos are far larger than the
dataset released in \cite{quality_action}. 

\vspace{0.1in}

\noindent \textbf{Scoring of figure skating}. We carefully annotated
each video with the skater and competition, and labeled it with two
scores, \emph{namely}, Total Element Score (TES) and Total Program
Component Score (PCS). These scores are given by the mark scheme of
figure skating competition. Specifically, these scores measure the
performance of skater at each stage over the whole competition. The
score of TES is used to judge the difficulty and execution of all
technical movement; and PCS aims at evaluating the performance and
interpretation of the music by the skaters. Both the TES and PCS are
given by nine different referees who are the experts on figure skating
competition.  Note that the same skater may receive very different
scores at different competition due to her performance. Finally we
gather 500 videos about ladies' singles short program, and each video
comes with the ground-truth scores. We randomly split the dataset
into 400 training videos and 100 testing ones.

\subsection{Data Analysis}

Apart from learning a score prediction model by using this dataset,
we conduct statistical analysis and have some interesting finding.
In particular, we compute the Spearman correlation and Kendall tau
correlation between TES and PCS over different matches (in Fig. \ref{fig:matches})
or different players (in Fig. \ref{fig:players}). More specific,
we take the TES and PCS values of all skaters in each match, and compute
the correlations as shown in Fig. \ref{fig:matches}. These values
reflect how the TES and PCS are correlated across different matches.
On the other hand, we take the same skater TES and PCs values of all
matches she took, and calculate their correlations in Fig. \ref{fig:players}.

As shown in Fig. \ref{fig:matches}, we find that in over a half of
all matches, the Toal Element Score (TES) has little correlation with
Total Program Component Score (PCS). This is reasonable, since the
TES and PCS are designed to measure two quite different perspectives
of the skater's performance in the whole competition. In other words,
TES and PCS should be relatively independent distributed. In a few
matches, we indeed observe the high correlation between TES and PCS
as in Fig. \ref{fig:matches}. We attribute this high correlation
to the subjectivity of referees, \emph{i.e}., referees would think
that the skaters who can complete difficult technical movements (TES)
are also able to interpret the music well (PCS). Furthermore, the
weak correlations between TES and PCS are also shown in Fig. \ref{fig:players}.

\noindent \textbf{Fis-V dataset Vs. MIT-skate dataset}. Comparing
with the existing MIT-skate dataset \cite{quality_action}, our dataset
has larger data scale (\emph{i.e.}, more than 3 times videos), higher
annotation quality (\emph{i.e.}, For each video, we provide both PCS
and TES scores provided, rather than a single total score), and collecting
more update-to-date figure skating videos (\emph{i.e.,} our videos
come from 12 competitions from 2012 to 2017) than the MIT-skate dataset.
Particularly, all of videos in MIT-Skate are from competitions happened
before 2012, which makes the dataset somehow outdated, since the scoring
standards of figure skating competitions is constantly changing in
the international competitions. We think a qualified figure skating
video dataset should be updated periodically.

\section{Methodology\label{sec:Methodology} }

In this section, we present our framework of learning to score the
figure skating videos. We divide the whole section into three parts.
Sec. \ref{subsec:Problem-Setup} discusses the problem setup and the
video features we are using. We discuss how to get video level representation
in Sec. \ref{subsec:Video-Level-Representation-by}. Finally, the
video fusion scheme of learning to score will be explained in Sec.
\ref{subsec:Video-fusion-by}. 

\begin{figure*}
\centering{}\includegraphics[scale=0.33]{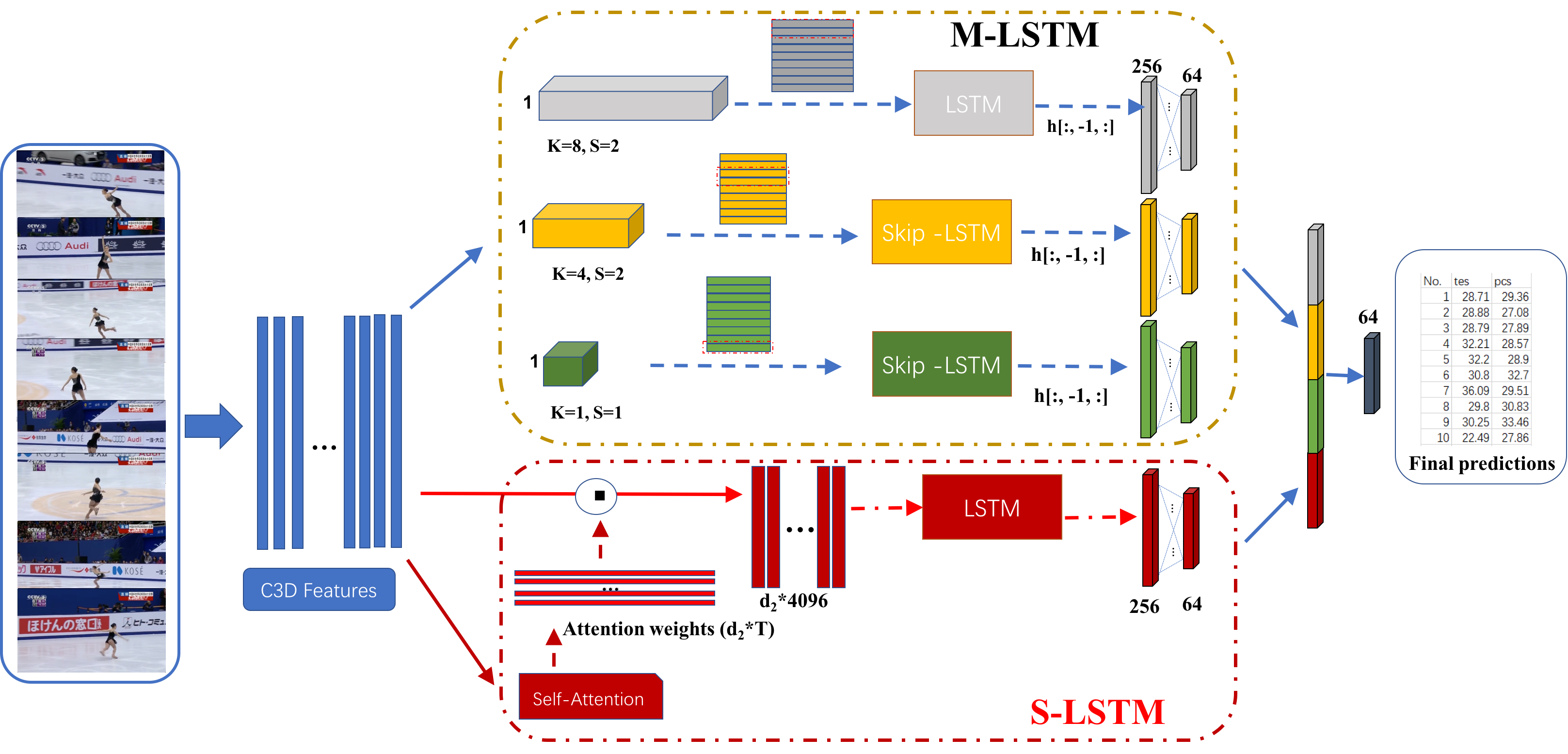}\caption{\label{fig:model}Overview of multi-scale convolution aggregation
model with skip-LSTM. }
\end{figure*}

\subsection{Problem Setup\label{subsec:Problem-Setup}}

\vspace{0.1in}

\noindent \textbf{Weakly labeled regression.} In figure skating matches,
the referees will incrementally add the TES with the progress of the
whole competition on-the-fly. Once the player finished one particular
technical movement, the corresponding TES and PCS scores will be added.
Ideally, we want the scores of each technical movement; but in the
real situation, it is impossible to get the incrementally added scores
synchronized with each video clip. Thus, we provide the final scores
of TES and PCS; and the tasks of predicting these scores can be formulated
as weakly labelled regression tasks. In our tasks, we take the prediction
of TES and PCS as two independent regression tasks.

\vspace{0.1in}

\noindent \textbf{Video Features. }We adopt deep spatial-temporal
convolution networks for more powerful video representation. We extract
deep clip-level features off-the-shelf from 3D Convolutional Networks,
which are pre-trained on large-scale dataset. In particular, We use
the 4096 dimensional clip-based feature from the \emph{fc6} layer
of \emph{C3D} \cite{Tran_ICCV2015} pre-trained on Sports-1M \cite{sports1m},
which is a large-scale dataset containing 1,133,158 videos which have
been annotated automatically with 487 sports labels. We use the sliding
window of size 16 frames over the video temporal to cut the video
clips with the stride as 8. 

\subsection{Self-Attentive LSTM (S-LSTM) \label{subsec:Video-Level-Representation-by}}

We propose a self-attentive feature embedding to selectively learn
to compact feature representations. Such representations can efficiently
model the local information. Specifically, since each video has about
4300 frames with 2 minutes and 50 seconds duration, the total computational
cost of using all C3D features would be very heavy. On the other hand,
the trivial practice is to employ max or average pooling operator
to merge these features into video-level representations. However,
not all video clips/frames contribute equally to regressing the final
scores. Thus in order to extract a more compact feature representation,
we have to address two problems properly,
\begin{enumerate}
\item The features of clips that are important to difficulty technical movements
should be heavy weighted.
\item The produced compact feature representations should be the fixed length
for all the videos.
\end{enumerate}
To this end, a self-attentive embedding scheme is proposed here to
generate the video-level representations. In particular, suppose we
have a $T\times d$ dimensional \emph{C3D} feature sequence of a video
$F=\left(f_{1},f_{2},\cdots,f_{T}\right)$, we can compute the weight
matrix $A$,
\begin{equation}
A=\sigma_{1}\left(W_{s2}\sigma_{2}\left(W_{s1}F^{T}\right)\right)\label{eq:attention}
\end{equation}

\noindent where $\sigma_{1}\left(\cdot\right)$ and $\sigma_{2}\left(\cdot\right)$
indicates the softmax and hyperbolic tangent function respectively.
The $\sigma_{1}\left(\cdot\right)$ can ensure the computed weights
sum to 1. 

We implement the Eq (\ref{eq:attention}) as a 2-layer Multiple Layer
Perceptron (MLP) without the bias of $d_{1}$ hidden neurons. Thus
the dimension of weights $W_{s1}$ and $W_{s2}$ are $d_{1}\times d$
and $d_{2}\times d_{1}$, and the dimension of $A$ is $d_{2}\times T$.
 The compact representation is computed as $M=A\cdot F$. Each row
of matrix $A$ can be interpreted as a specific focus point on the
video, maybe a key action pattern; the $d_{1}$ stands for the diversity
of descriptions. Therefore multiplying feature matrix $F$ with $A$
helps us extract all such patterns, resulting in a shorter input sequence,
with dimension of $d_{2}\times d$. 

The resulting embedding $M$ is further followed by a 1-layer Long
Short-Term Memory (LSTM) with the $d_{2}$ LSTM cell. The output LSTM
is further connected to a 1-layer fully connected layer with 64 neurons
to regress the TES and PCS scores. We use the Mean Square Error (MSE)
as the loss function to optimize this self-attentive LSTM. A penalty
term is added to the MSE loss function in order to encourage the diversity
of learned self-attentive feature embedding $M$. The form of the
penalty is,
\begin{equation}
P=\left\Vert \left(AA^{T}-I\right)\right\Vert _{F}^{2}\label{eq:p}
\end{equation}

\noindent where $\left\Vert \cdot\right\Vert _{F}^{2}$ is the Frobenius
norm. $I$ is the identity matrix. 

\begin{figure}
\centering{}\includegraphics[scale=0.46]{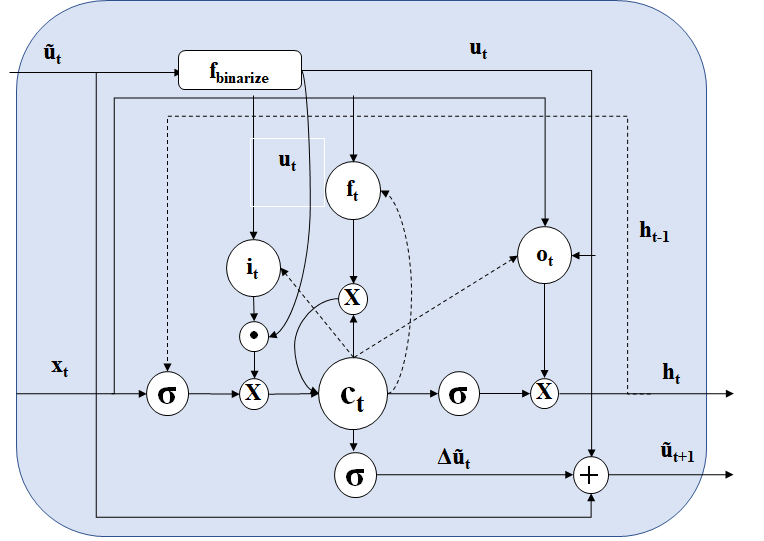}\caption{\label{fig:skip}Structure of revised skip-LSTM cell. }
\end{figure}

The self-attentive LSTM is for the first time proposed here to address
the regression tasks. We highlight several differences with previous
works. (1) The attention strategy has been widely utilized in previous
works \cite{yunansigir2017,show2015tell,li2016videolstm,sermanet2014attention}.
In contrast, the self-attentive strategy simply uses the final output
of video sequences. Similar strategy has also been used in the NLP
tasks \cite{lin2017structured}. (2) Comparing with \cite{lin2017structured},
our self-attentive LSTM is also very different. The output of self-attentive
feature embedding is used as the input of LSTM and fully connected
layer for the regression tasks. In contrast, \cite{lin2017structured}
utilized the attention strategy to process the output of LSTM, and
directly concatenate the feature embedding for the classification
tasks. 

\subsection{Multi-scale Convolutional Skip LSTM (M-LSTM) \label{subsec:Video-fusion-by}}

The self-attentive LSTM is efficient in modeling the local sequential
information. Nevertheless, it is essential to model the sequential
frames/clips containing the local (technical movements) and global
(performance of players), since in principle, TES scores the technical
movement, and the PCS reflects the whole performance of the player.
 In light of this understanding, we propose the multi-scale convolutional
skip LSTM (M-LSTM) model. 

As an extension of LSTM, our M-LSTM learns to model the sequential
information at multiple scale. Specifically, the dense clip-based
C3D video features give the good representations of local sequential
information. To facilitate abstracting the information of multiple
scale, our M-LSTM employs several parallel $1$D convolution layers
with different kernel sizes, as shown in Fig. \ref{fig:model}. 
The kernel with small size of filters can aggregate and extract the
visual representation of action patterns lasting seconds in the videos.
The kernel of large size of filters will try to model the global information
of the videos. However, in practice, quite different from the videos
used for video classification (e.g., UCF101 \cite{Soomro2012}), our
figure skating videos are quite longer. Thus the total frames of our
figure staking videos still make the training process of LSTM difficulty
in capturing long term dependencies. 

To solve this issue, we further propose the skipping RNN strategy
here. Particularly, we propose the revised skip LSTM structure. An
origin LSTM works as follows:
\begin{align}
i_{t},f_{t},o_{t}= & \sigma\left(W_{x}x_{t}+W_{h}h_{t-1}+b\right)\label{eq:ifo}\\
g_{t}= & tanh\left(W_{xg}x_{t}+W_{hg}h_{t-1}+b_{g}\right)\label{eq:g}\\
c_{t}= & f_{t}\odot c_{t-1}+i_{t}\odot g_{t}\label{eq:c}\\
h_{t}= & o_{t}\odot tanh\left(c_{t}\right)\label{eq:h}
\end{align}

\noindent where $i_{t},f_{t},o_{t}$ are the input, forget and output
gates. $\sigma\left(\cdot\right)$ indicates sigmod function. $x_{t}$
is the input of LSTM; the hidden state and cell state of LSTM are
denoted as $h_{t}$ and $c_{t}$ respectively. $W_{x},W_{h}$ $W_{xg},W_{hg}$
and $b,b_{g}$ are the learning weights of parameters. In skip LSTM,
a binary state update gate, $u_{t}\in\left\{ 0,1\right\} $ is added,
which is used to control the update of cell state and hidden state.
The whole new update rule is as follows,
\begin{align}
u_{t} & =f_{binary}\left(\tilde{u}_{t}\right)\label{eq:u_skip}\\
i_{t},f_{t},o_{t} & =\sigma\left(W_{x}x_{t}+W_{h}h_{t-1}+b\right)\\
g_{t} & =tanh\left(W_{xg}x_{t}+W_{hg}h_{t-1}+b_{g}\right)\\
\tilde{c_{t}} & =f_{t}\odot c_{t-1}+i_{t}\odot g_{t}\\
\tilde{h_{t}} & =o_{t}\odot tanh\left(c_{t}\right)\\
c_{t} & =u_{t}\cdot\tilde{c}_{t}+\left(1-u_{t}\right)\cdot c_{t-1}\\
h_{t} & =u_{t}\cdot\tilde{h}_{t}+\left(1-u_{t}\right)\cdot h_{t-1}\\
\Delta\tilde{u}_{t} & =\sigma\left(W_{p}c_{t}+b_{p}\right)\\
\tilde{u}_{t+1} & =u_{t}\cdot\Delta\tilde{u}_{t}+\left(1-u_{t}\right)\cdot\left(\tilde{u}_{t}+\min\left(\Delta\tilde{u}_{t},1-\tilde{u}_{t}\right)\right)
\end{align}

\noindent Where $\sigma\left(\cdot\right)$ is sigmoid function,
$\odot$ denotes element-wise multiplication. $f_{binary}\left(\cdot\right)$
indicates the round function. $\tilde{c}_{t}$ and $\tilde{h_{t}}$
are the values of corresponding state $c_{t}$ and $h_{t}$ if $u_{t}=1$.
$\Delta\tilde{u}_{t}$ is the accumulated error if not updating the
control variable $u_{t}$. Furthermore, different from \cite{campos2017skip},
our model revises the update rule of $c_{t}$ and $h_{t}$ to prevent
the network from being forced to expose a memory cell which has not
been updated, which would result in misleading information, as shown
in Fig. \ref{fig:skip}.
\begin{align}
c_{t} & =f_{t}\odot c_{t-1}+u_{t}\cdot i_{t}\odot g_{t}\label{eq:ct}\\
h_{t} & =\left(\left(1-u_{t}\right)\cdot o_{t}+u_{t}\cdot o_{t-1}\right)\odot tanh\left(c_{t}\right)\label{eq:ht}
\end{align}

The key ingredient of our skip LSTM lies in Eq (\ref{eq:u_skip}).
By using the round function, our M-LSTM can skip some less significant
update if $u_{t}=0$. By virtue of this way, our M-LSTM can model
even longer term data dependencies. 

The whole structure of our M-LSTM is also illustrated in Fig. \ref{fig:model}.
Since the skip LSTM is used to discard redundant information, we only
connect it to the convolution layers with small-size kernels, and
apply the common LSTM after other convolution layers. The outputs
at the final time-step of all parallel LSTMs are then concatenated
and transmitted to a fully connected layer to regress the prediction
scores. \textcolor{red}{ }

Thus, with this M-LSTM architecture, we can actually have the best
of both world: the multi-scale convolutional structures can extract
the local and global feature representations from videos; the revised
skip LSTM can efficiently skip/discard the redundant information that
is not essential in learning the local and global information. The
final LSTM outputs of different scales are still concatenated and
learned by the nonlinear fully connected layer for the regression.
The effectiveness of our M-LSTM is validated in the experiments.

\section{Experiments\label{sec:Experiments}}

\subsection{Settings and Evaluation}

\noindent \textbf{Datasets. }We evaluate our tasks in both MIT-skate
\cite{quality_action} and our Fis-V dataset. MIT-skate has 150 videos
with 24 frames per second. We utilize the standard data split of 100
videos for training and the rest for testing. In our Fis-V dataset,
we introduce the split of 400 videos as training, the rest as testing. 

\vspace{0.1in}

\noindent \textbf{Metric Settings}. As for the evaluation, we use
the standard evaluation metrics, the spearman correlation \textcolor{red}{$\rho$
}\textendash{} proposed in \cite{quality_action} and \cite{parmar2017learning}.
This makes the results of our framework directly comparable to those
results reported in \cite{quality_action,parmar2017learning}. Additionally,
to give more insights of our model, the Mean Square Error (MSE) is
also utilized here to evaluate the models. In MIT-skate, the published
results are trained on the final scores; so these scores have been
used to evaluate our framework.

\vspace{0.1in}

\noindent \textbf{Experimental Settings}. For self-attention LSTM
subnetwork, we set $d_{1}=1024,d_{2}=40$, and the hidden size of
LSTM is also set as $256$. The batch size is 32. For M-LSTM subnetwork,
we use the hidden size of $256$ for both types of LSTM layers, the
other parameter setting is depicted in Fig. \ref{fig:model}. For
both models we use a two-layer perceptron with a hidden size of 256
and ReLU function as the activation function of hidden layer. We build
both models by Pytorch and optimize the model by Adam \cite{kingma2014adam}
algorithm with learning rate of $1e-4$. The whole framework is trained
on 1 NVIDIA 1080Ti GPU card and can get converged by 250 epochs. It
totally takes 20 minutes to train one model. We augment the videos
by the horizontal flipping on frames. As the standard practice, the
Dropout is set as 0.7 and only used in fully connected layers; batch
normalization is added after each convolution layer in our model.
Our model is an end-to-end network; so, we directly use the C3D feature
sequences of training data to train the model with the parameters
above. 

Addtitionally, we donot fine-tune the C3D features in our model, due
to the tremendous computational cost. Literally, our videos are very
long (averagely 4400 frames), but only around 400 videos. So if we
want to finetune C3D with Fis-V or MIT-skate dataset, we need to forward
pass and backpropagate on this relatively large C3D model 400 iteration
for each video. This requires huge computational cost. On the other
hand, we have observed overfitting in our training if the hyperparameter
is not well tuned, due to the small dataset size (only 400 videos).
Thus adding C3D into training graph would make the training process
more difficult. 

\vspace{0.1in}

\noindent \textbf{Competitors.} Several different competitors and
variants are discussed here. Specifically, we consider different combinations
of the following choices:
\begin{enumerate}
\item Using frame-level features: We use the 2048 dimensional feature from
the pool5 layer of the SENet \cite{SEnet}, which is the winner of
ILSVRC 2017 Image Classification Challenge.
\item Using max or average pooling for video-level representation. 
\item Using different regression models: SVR with linear or RBF kernels
are utilized for regression tasks. 
\item LSTM and bi-LSTM based models. We use the C3D-LSTM model depicted
in \cite{parmar2017learning}. Note that due to very long video sequence
and to make a more fair comparison, we set the hidden size of LSTM
as $256$, adopt an option of bi-directional LSTM, and use a multi-layer
regressor same as our models. This C3D-LSTM model is extended to using
SENet features, or by using bi-directional LSTM. 
\item \cite{quality_action,hierarchical}. We also report the results of
these two papers. 
\end{enumerate}

\subsection{Results}

\begin{table}
\begin{centering}
\begin{tabular}{c|c|c|c|c|c}
\hline 
\multirow{2}{*}{Features} & \multirow{2}{*}{Pooling} & \multirow{2}{*}{Reg} & \multirow{2}{*}{MIT-skate} & \multicolumn{2}{c}{Fis-V}\tabularnewline
\cline{5-6} 
 &  &  &  & TES & PCS \tabularnewline
\hline 
\hline 
\cite{quality_action} & \multicolumn{2}{c|}{\textemdash{}} & 0.33 & \textendash{} & \textendash{}\tabularnewline
\hline 
\cite{hierarchical} & \multicolumn{2}{c|}{\textemdash{}} & 0.45 & \textendash{} & \textendash{}\tabularnewline
\hline 
\cite{parmar2017learning} & \multicolumn{2}{c|}{\textemdash{}} & 0.53 & \textendash{} & \textendash{}\tabularnewline
\hline 
\hline 
\multirow{6}{*}{SENet} & Max & Linear & \textendash{} & 0.39 & 0.53\tabularnewline
\cline{2-6} 
 & Avg & Linear & \textendash{} & 0.43 & 0.61\tabularnewline
\cline{2-6} 
 & Max & RBF & \textendash{} & 0.27 & 0.43\tabularnewline
\cline{2-6} 
 & Avg & RBF & \textendash{} & 0.21 & 0.34\tabularnewline
\cline{2-6} 
 & \multicolumn{2}{c|}{LSTM} & \textendash{} & 0.57 & 0.70\tabularnewline
\cline{2-6} 
 & \multicolumn{2}{c|}{bi-LSTM} & \textendash{} & 0.57 & 0.70\tabularnewline
\hline 
\multirow{6}{*}{C3D} & Max & Linear & 0.48 & 0.47 & 0.61\tabularnewline
\cline{2-6} 
 & Avg & Linear & 0.40 & 0.40 & 0.59\tabularnewline
\cline{2-6} 
 & Max & RBF & 0.44 & 0.35 & 0.49\tabularnewline
\cline{2-6} 
 & Avg & RBF & 0.42 & 0.41 & 0.56\tabularnewline
\cline{2-6} 
 & \multicolumn{2}{c|}{LSTM} & 0.37 & 0.59 & 0.77\tabularnewline
\cline{2-6} 
 & \multicolumn{2}{c|}{bi-LSTM} & 0.58 & 0.56 & 0.73\tabularnewline
\hline 
\hline 
\multirow{3}{*}{C3D} & \multicolumn{2}{c|}{M-LSTM} & 0.56 & 0.65 & 0.78\tabularnewline
\cline{2-6} 
 & \multicolumn{2}{c|}{S-LSTM} & 0.51 & \textbf{0.67} & 0.77\tabularnewline
\cline{2-6} 
 & \multicolumn{2}{c|}{S-LSTM+M-LSTM} & \textbf{0.59} & 0.65 & \textbf{0.78}\tabularnewline
\hline 
\end{tabular}
\par\end{centering}
\caption{\label{table:regression}Results of the spearman correlation (the
higher the better) on MIT-Skate and Fis-V. ``S-LSTM'' is short for
self-attentive LSTM. ``M-LSTM'' is short for Multi-scale Convolutional
Skip LSTM used.}
\end{table}
\begin{table}
\begin{centering}
\begin{tabular}{c|c|c|c|c}
\hline 
\multirow{1}{*}{Features		} & \multirow{1}{*}{Pooling} & \multirow{1}{*}{Reg} & TES & PCS\tabularnewline
\hline 
\hline 
\multirow{6}{*}{SENet} & Max & Linear & 32.35 & 16.14\tabularnewline
\cline{2-5} 
 & Avg & Linear & 34.15 & 15.15\tabularnewline
\cline{2-5} 
 & Max & RBF & 37.78 & 22.07\tabularnewline
\cline{2-5} 
 & Ag & RBF & 39.69 & 24.70\tabularnewline
\cline{2-5} 
 & \multicolumn{2}{c|}{LSTM} & 23.38 & 11.03\tabularnewline
\cline{2-5} 
 & \multicolumn{2}{c|}{bi-LSTM} & 24.65 & 11.19\tabularnewline
\hline 
\multirow{6}{*}{C3D} & Max & Linear & 27.42 & 13.98\tabularnewline
\cline{2-5} 
 & Avg & Linear & 30.25 & 15.96\tabularnewline
\cline{2-5} 
 & Max & RBF & 40.19 & 25.13\tabularnewline
\cline{2-5} 
 & Avg & RBF & 34.60 & 19.08\tabularnewline
\cline{2-5} 
 & \multicolumn{2}{c|}{LSTM} & 22.96 & 8.70\tabularnewline
\cline{2-5} 
 & \multicolumn{2}{c|}{bi-LSTM} & 23.80 & 10.36\tabularnewline
\hline 
\hline 
\multirow{3}{*}{C3D} & \multicolumn{2}{c|}{M-LSTM} & 19.49 & 8.41\tabularnewline
\cline{2-5} 
 & \multicolumn{2}{c|}{S-LSTM} & \textbf{19.26} & 8.53\tabularnewline
\cline{2-5} 
 & \multicolumn{2}{c|}{S-LSTM+M-LSTM} & 19.91 & \textbf{8.35}\tabularnewline
\hline 
\end{tabular}
\par\end{centering}
\caption{\label{table:regression-1}Results of the MSE on Fis-V. }
\end{table}

\begin{figure*}
\begin{centering}
\includegraphics[scale=0.35]{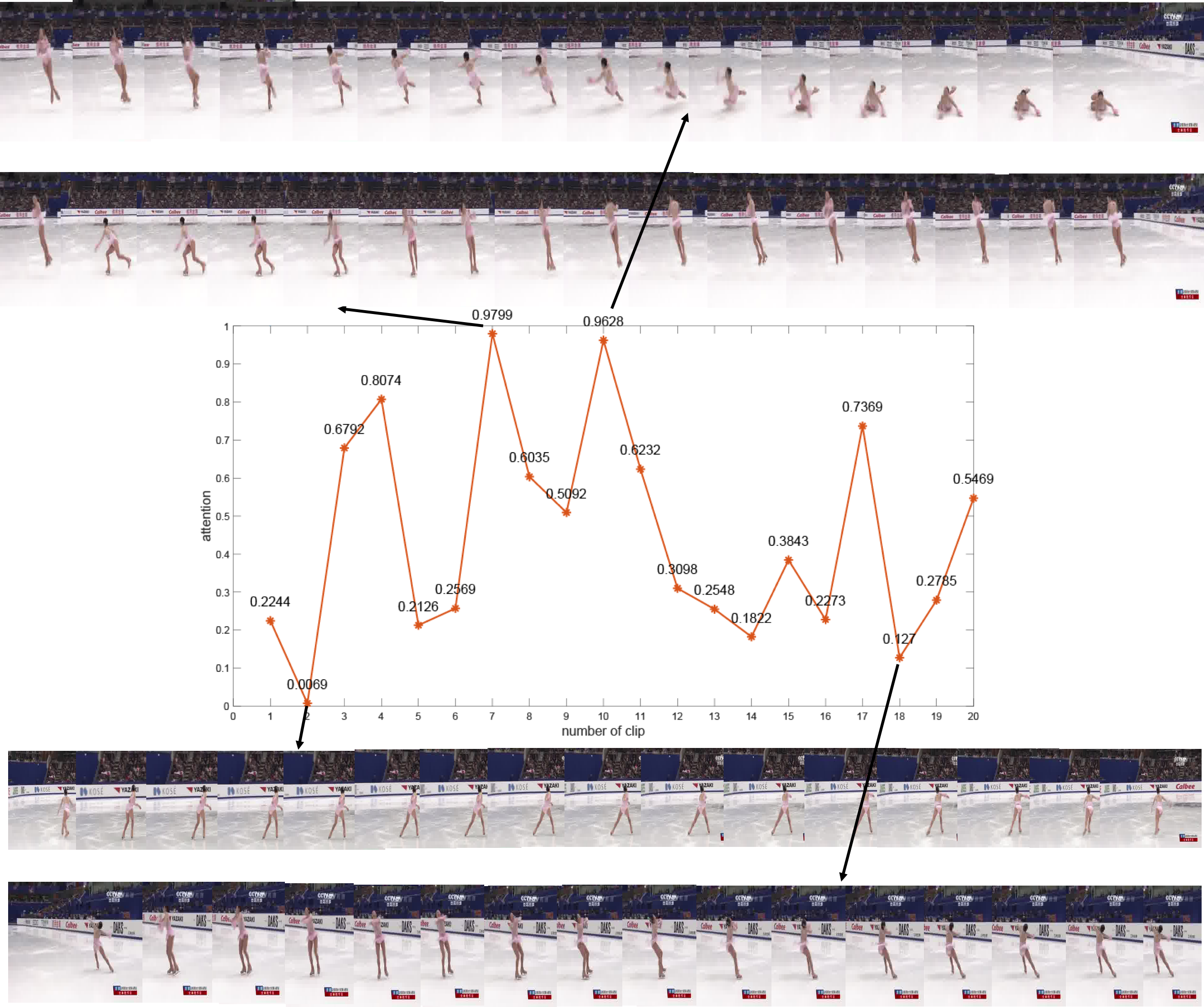}
\par\end{centering}
\caption{\label{fig:visualize}Qualitative study on self-attention model. The
top two rows show a clip with high attention weights while the bottom
two rows show clips of low attention weights. }
\end{figure*}

\vspace{0.1in}

\noindent \textbf{Results of the spearman correlation.} We report
the results in Tab. \ref{table:regression}. On MIT-skate and our
Fis-V dataset, we compare several variants and baselines. We highlight
that our framework achieves the best performance on both datasets,
and outperform the baselines (including \cite{quality_action,hierarchical,parmar2017learning})
clearly by a large margin. This shows the effectiveness of our proposed
framework. We further conduct the ablation study to explore the contributions
of each components, namely, M-LSTM and S-LSTM. In general, the results
of M-LSTM can already beat all the other baselines on both datasets.
This is reasonable, since the M-LSTM can effectively learn the local
and global information with the efficient revised skip LSTM structure.
Further, on MIT-skate dataset the S-LSTM is complementary to M-LSTM,
since we can achieve higher results. The performance of M-LSTM and
S-LSTM is very good on Fis-V dataset. 

\vspace{0.1in}

\noindent \textbf{Results of different variants.} As the regression
tasks, we further explore different variants in Tab. \ref{table:regression}.
By using the C3D features, we compare different pooling and Regression
methods. 

(1) \emph{Max Vs. Avg} \emph{pooling}. Actually, we donot have conclusive
results which pooling method is better. The max pooling has better
performance than average pooling on MIT-skate dataset, while the average
pooling can beat the maximum pooling on Fis-V dataset. This shows
the difficult intrinsic of the regression tasks. 

(2) \emph{RBF Vs. Linear SVR}. In general, we found that the linear
SVR has better performance than the RBF SVR. And both methods have
lower performance than our framework. 

(3) \emph{SENet Vs. C3D.} On Fis-V dataset, we also compare the results
of using SENet features. SENet are the static frame-based features,
and C3D are clip-based features. Note that within each video, we generally
extract different number of SENet and C3D features; thus it is nontrivial
to directly combine two types of features together. Also the models
using C3D features can produce better prediction results than those
from SENet, since the figure skating videos are mostly about the movement
of each skater. The clip-based C3D features can better abstract this
moving information from the videos. 

(4) \emph{TES Vs. PCS}. With comparable models and features, the correlation
results on PCS are generally better than those of TES. This reflects
that the PCS is relatively easier to be predicted than TES. 

\vspace{0.1in}

\noindent \textbf{Results of the mean square error.} On our Fis-V
dataset, we also compare the results by the metrics of MSE in Tab.
\ref{table:regression-1}. In particular, we find that the proposed
M-LSTM and S-LSTM can significantly beat all the other baseline clearly
by a large margin. Furthermore, we still observe a boosting of the
performance of PCS by combining the M-LSTM and S-LSTM. 

Interestingly, we notice that on TES, the combination of S-LSTM+M-LSTM
doesnot have significantly improved over the M-LSTM or S-LSTM only.
This is somehow expected. Since the TES task aims at scoring those
clips of technical movements. This is relatively a much easier task
than the PCS task which aims at scoring the good interpretation of
music. Thus only features extracted by M-LSTM or S-LSTM can be good
enough to learn a classifier for TES. The combination of both M-LSTM
and S-LSTM may lead to redundant information. Thus the S-LSTM+M-LSTM
can not get further improvement over M-LSTM or S-LSTM on TES.

\vspace{0.1in}

\noindent \textbf{Ablation study on Self-Attentive strategy. }To
visualize the self-attentive mechanism, we compute the attention weight
matrix $A$ of a specific video. We think that if a clip has high
weight in at least one row of $A$, then it means that this clip shows
an important technical movement contributing to the TES score, otherwise
it is insignificant. We show a pair of example clips (16 frames) in
Fig. \ref{fig:visualize}. From Fig. \ref{fig:visualize} we can find
the clip in the top two rows with higher attention weights is showing
``hard'' action, for instance, ``jumping on the same foot within
a spin'', while movement in the bottom clips would not.\textcolor{red}{{}
}

\section{Conclusion\label{sec:Conclusion}}

In this paper, we present a new dataset \textendash{} Fis-V dataset
for figure skating sports video analysis. We target the task of learning
to score of each skater's performance. We propose two models for the
regression tasks, \emph{namely}, the Self-Attentive LSTM (S-LSTM)
and the Multi-scale Convolutional Skip LSTM (M-LSTM). We also integrate
the two proposed networks in a single end-to-end framework. We conduct
extensive experiments to thoroughly evaluate our frameworks as well
as the variants on MIT-skate and Fis-V dataset. The experimental results
validate the effectiveness of proposed methods. 

\bibliographystyle{plain}
\bibliography{emotion}

\begin{thebibliography}{10}

\bibitem{mkl_2004}
F.~R. Bach, G.~R. Lanckriet, and M.~I. Jordan.
\newblock Multiple kernel learning, conic duality, and the smo algorithm.
\newblock In {\em ICML}, 2004.

\bibitem{bilen2016dynamic}
Hakan Bilen, Basura Fernando, Efstratios Gavves, Andrea Vedaldi, and Stephen
  Gould.
\newblock Dynamic image networks for action recognition.
\newblock In {\em CVPR}, 2016.

\bibitem{campos2017skip}
V{\'\i}ctor Campos, Brendan Jou, Xavier Gir{\'o}-i Nieto, Jordi Torres, and
  Shih-Fu Chang.
\newblock Skip rnn: Learning to skip state updates in recurrent neural
  networks.
\newblock {\em ICLR}, 2018.

\bibitem{heterog_iccv_2009}
L.~Cao, J.~Luo, F.~Liang, and T.~S. Huang.
\newblock Heterogeneous feature machines for visual recognition.
\newblock In {\em ICCV}, 2009.

\bibitem{MBH}
N.~Dalal, B.~Triggs, and C.~Schmid.
\newblock Human detection using oriented histograms of flow and appearance.
\newblock In {\em ECCV}, 2006.

\bibitem{person_interaction}
V.~Delaitre, J.~Sivic, and I.~Laptev.
\newblock Learning person-object interactions for action recognition in still
  images.
\newblock In {\em NIPS}, 2011.

\bibitem{donahue_cvpr2015}
J.~Donahue, L.~Anne Hendricks, S.~Guadarrama, M.~Rohrbach, S.~Venugopalan,
  K.~Saenko, and T.~Darrell.
\newblock Long-term recurrent convolutional networks for visual recognition and
  description.
\newblock In {\em CVPR}, 2015.

\bibitem{efros2003action}
A.~A. Efros, A.~C. Berg, G.~Mori, and J.~Malik.
\newblock Recognizing action at a distance.
\newblock In {\em IEEE International Conference on Computer Vision}, pages
  726--733, 2003.

\bibitem{Feichtenhofer16}
C.~Feichtenhofer, A.~Pinz, and A.~Zisserman.
\newblock Convolutional two-stream network fusion for video action recognition.
\newblock In {\em CVPR}, 2016.

\bibitem{gordon_1995}
A.S Gordon.
\newblock Automated video assessment of human performance.
\newblock In {\em AI-ED}, 1995.

\bibitem{gupta_pami2009}
Gupta, Kembhavi A., and L.S Davis.
\newblock Observing human-object interactions: Using spatial and functional
  compatibility for recognitions.
\newblock In {\em IEEE TPAMI}, 2009.

\bibitem{taylor_eccv2010}
G.W.Taylor, R.Fergus, Y.LeCun, and C.Bregler.
\newblock Convolutional learning of spatio-temporal features.
\newblock In {\em ECCV}, 2010.

\bibitem{activitynet}
Fabian~Caba Heilbron, Victor Escorcia, Bernard Ghanem, and Juan~Carlos Niebles.
\newblock Activitynet: A large-scale video benchmark for human activity
  understanding.
\newblock In {\em CVPR}, 2015.

\bibitem{SEnet}
Jie Hu, Li~Shen, and Gang Sun.
\newblock Squeeze-and-excitation networks.
\newblock In {\em arxiv}, 2017.

\bibitem{dense_traj}
H.Wang, A.~Kl\"aser, C.~Schmid, and C.-L. Liu.
\newblock Action recognition by dense trajectories.
\newblock In {\em CVPR}, 2011.

\bibitem{Ji2010}
Shuiwang Ji, Wei Xu, Ming Yang, and Kai Yu.
\newblock 3d convolutional neural networks for human action recognition.
\newblock In {\em ICML}, 2010.

\bibitem{jiang_acmmm2009}
W.~Jiang, C.~Cotton, S.-F. Chang, D.~Ellis, and A.~Loui.
\newblock Short-term audio-visual atoms for generic video concept
  classification.
\newblock In {\em ACM MM}, 2009.

\bibitem{fcvid_2017}
Yu-Gang Jiang, Zuxuan Wu, Jun Wang, Xiangyang Xue, and Shih-Fu Chang.
\newblock Exploiting feature and class relationships in video categorization
  with regularized deep neural networks.
\newblock In {\em IEEE TPAMI}, 2017.

\bibitem{jiang2011consumervideo}
Yu-Gang Jiang, Guangnan Ye, Shih-Fu Chang, Daniel Ellis, and Alexander~C. Loui.
\newblock Consumer video understanding: A benchmark database and an evaluation
  of human and machine performance.
\newblock In {\em ACM International Conference on Multimedia Retrieval}, 2011.

\bibitem{jug_2003}
Marko Jug, Janez Per{\v{s}}, Branko De{\v{z}}man, and Stanislav
  Kova{\v{c}}i{\v{c}}.
\newblock Trajectory based assessment of coordinated human activity.
\newblock In {\em International Conference on Computer Vision Systems}, pages
  534--543. Springer, 2003.

\bibitem{deepvideo}
Andrej Karpathy, George Toderici, Sanketh Shetty, Thomas Leung, Rahul
  Sukthankar, and Li~Fei-Fei.
\newblock Large-scale video classification with convolutional neural networks.
\newblock In {\em CVPR}, 2014.

\bibitem{sports1m}
Andrej Karpathy, George Toderici, Sanketh Shetty, Thomas Leung, Rahul
  Sukthankar, and Li~Fei-Fei.
\newblock Large-scale video classification with convolutional neural networks.
\newblock In {\em CVPR}, 2014.

\bibitem{kingma2014adam}
Diederik Kingma and Jimmy Ba.
\newblock Adam: A method for stochastic optimization.
\newblock In {\em ICLR}, 2015.

\bibitem{KrizhevskySH12}
Alex Krizhevsky, Ilya Sutskever, and Geoffrey~E. Hinton.
\newblock Imagenet classification with deep convolutional neural networks.
\newblock In {\em NIPS}, 2012.

\bibitem{HMDB51}
H.~Kuehne, H.~Jhuang, E.~Garrote, T.~Poggio, and T.~Serre.
\newblock {HMDB}: A large video database for human motion recognition.
\newblock In {\em ICCV}, 2011.

\bibitem{laptev2008actions}
I.~Laptev, M.~Marszalek, C.~Schmid, and B.~Rozenfeld.
\newblock Learning realistic human actions from movies.
\newblock In {\em IEEE Conference on Computer Vision and Pattern Recognition},
  pages 1--8, 2008.

\bibitem{retrieving_movie}
I.~Laptev and P.~Perez.
\newblock Retrieving actions in movies.
\newblock In {\em ICCV}, 2007.

\bibitem{hierarchical}
Quoc~V Le, Will~Y Zou, Serena~Y Yeung, and Andrew~Y Ng.
\newblock Learning hierarchical invariant spatio-temporal features for action
  recognition with independent subspace analysis.
\newblock In {\em Computer Vision and Pattern Recognition (CVPR), 2011 IEEE
  Conference on}, pages 3361--3368. IEEE, 2011.

\bibitem{li2016videolstm}
Zhenyang Li, Efstratios Gavves, Mihir Jain, and Cees~GM Snoek.
\newblock Videolstm convolves, attends and flows for action recognition.
\newblock {\em arXiv preprint arXiv:1607.01794}, 2016.

\bibitem{lin2017structured}
Zhouhan Lin, Minwei Feng, Cicero Nogueira~dos Santos, Mo~Yu, Bing Xiang, Bowen
  Zhou, and Yoshua Bengio.
\newblock A structured self-attentive sentence embedding.
\newblock {\em arXiv preprint arXiv:1703.03130}, 2017.

\bibitem{liu_2013_cvpr}
D.~Liu, K.-T. Lai, G.~Ye, M.-S. Chen, and S.-F. Chang.
\newblock Sample-specific late fusion for visual category recognition.
\newblock In {\em CVPR}, 2013.

\bibitem{camera_light}
Zach Lowe.
\newblock Lights, cameras, revolution.
\newblock {\em Grantland, March}, 2013.

\bibitem{distributed_repre_cvpr2011}
S.~Maji, L.~Bourdev, and J.~Malik.
\newblock Action recognition from a distributed representation of pose and
  appearance.
\newblock In {\em CVPR}, 2011.

\bibitem{McQueen2014}
A.~McQueen, J.~Wiens, and J.~Guttag.
\newblock Automatically recognizing on-ball screens.
\newblock In {\em MIT Sloan Sports Analytics Conference (SSAC)}, 2014.

\bibitem{natarajan_2012}
P.~Natarajan, S.~Wu, S.~Vitaladevuni, X.~Zhuang, S.~Tsakalidis, U.~Park, and
  R.~Prasad.
\newblock Multimodal feature fusion for robust event detection in web videos.
\newblock In {\em CVPR}, 2012.

\bibitem{snippet_cvpr2015}
J.~Yue-Hei Ng, M.~Hausknecht, S.~Vijayanarasimhan, O.~Vinyals, R.~Monga, and
  G.~Toderici.
\newblock Beyond short snippets: Deep networks for video classification.
\newblock In {\em CVPR}, 2015.

\bibitem{multimodal_icml2011}
J.~Ngiam, A.~Khosla, M.~Kim, J.~Nam, H.~Lee, and A.~Ng.
\newblock Multimodal deep learning.
\newblock In {\em ICML}, 2011.

\bibitem{over2011trecvid}
P.~Over, G.~Awad, M.~Michel, J.~Fiscus, W.~Kraaij, and A.~F. Smeaton.
\newblock Trecvid 2011 -- an overview of the goals, tasks, data, evaluation
  mechanisms and metrics.
\newblock In {\em Proceedings of TRECVID 2011}, 2011.

\bibitem{parmar2017learning}
Paritosh Parmar and Brendan~Tran Morris.
\newblock Learning to score olympic events.
\newblock In {\em Computer Vision and Pattern Recognition Workshops (CVPRW),
  2017 IEEE Conference on}, pages 76--84. IEEE, 2017.

\bibitem{perse_2007}
M.~Perse, M.~Kristan, J.~Pers, and S~Kovacic.
\newblock Automatic evaluation of organized basketball activity using bayesian
  networks.
\newblock In {\em Citeseer}, 2007.

\bibitem{quality_action}
H.~Pirsiavash, C.~Vondrick, and Torralba.
\newblock Assessing the quality of actions.
\newblock In {\em ECCV}, 2014.

\bibitem{two_strems_nips}
M.~Yang S.~Ji, W.~Xu and K.~Yu.
\newblock Convolutional two-stream network fusion for video action recognition.
\newblock In {\em NIPS}, 2014.

\bibitem{action_bank}
S.~Sadanand and J.J. Corso.
\newblock Action bank: A high-level representation of activity in video.
\newblock In {\em CVPR}, 2012.

\bibitem{sermanet2014attention}
Pierre Sermanet, Andrea Frome, and Esteban Real.
\newblock Attention for fine-grained categorization.
\newblock {\em arXiv}, 2014.

\bibitem{show2015tell}
Attend Show.
\newblock Tell: Neural image caption generation with visual attention.
\newblock {\em Kelvin Xu et. al.. arXiv Pre-Print}, 2015.

\bibitem{Simonyan2014}
Karen Simonyan and Andrew Zisserman.
\newblock Two-stream convolutional networks for action recognition in videos.
\newblock In {\em NIPS}, 2014.

\bibitem{Soomro2012}
Khurram Soomro, Amir~Roshan Zamir, and Mubarak Shah.
\newblock {UCF101:} {A} dataset of 101 human actions classes from videos in the
  wild.
\newblock {\em CRCV-TR-12-01}, 2012.

\bibitem{Srivastava_NIPS2012}
N.~Srivastava and R.~Salakhutdinov.
\newblock Multimodal learning with deep boltzmann machines.
\newblock In {\em NIPS}, 2012.

\bibitem{sun2015human}
Lin Sun, Kui Jia, Dit-Yan Yeung, and Bertram~E Shi.
\newblock Human action recognition using factorized spatio-temporal
  convolutional networks.
\newblock In {\em CVPR}, 2015.

\bibitem{Tran_ICCV2015}
D.~Tran, L.~Bourdev, R.~Fergus, L.~Torresani, and M.~Paluri.
\newblock Learning spatiotemporal features with 3d convolutional networks.
\newblock In {\em ICCV}, 2015.

\bibitem{C3D}
Du~Tran, Lubomir~D Bourdev, Rob Fergus, Lorenzo Torresani, and Manohar Paluri.
\newblock C3d: Generic features for video analysis.
\newblock In {\em ICCV}, 2015.

\bibitem{improved_dense}
H.~Wang and C.~Schmid.
\newblock Action recognition with improved trajectories.
\newblock In {\em ICCV}, 2013.

\bibitem{Wang2013a}
Heng Wang and Cordelia Schmid.
\newblock Action recognition with improved trajectories.
\newblock In {\em ICCV}, 2013.

\bibitem{cvpr15:wang}
L.~Wang, Y.~Qiao, and X.~Tang.
\newblock Action recognition with trajectory-pooled deep-convolutional
  descriptors.
\newblock In {\em CVPR}, 2015.

\bibitem{WangXWQLTV16}
Limin Wang, Yuanjun Xiong, Zhe Wang, Yu~Qiao, Dahua Lin, Xiaoou Tang, and Luc
  {Van Gool}.
\newblock Temporal segment networks: Towards good practices for deep action
  recognition.
\newblock In {\em ECCV}, 2016.

\bibitem{Wang_Transformation}
Xiaolong Wang, Ali Farhadi, and Abhinav Gupta.
\newblock Actions {\textasciitilde} transformations.
\newblock In {\em CVPR}, 2016.

\bibitem{pyarmid_video_fusion}
Yunbo Wang, Mingsheng Long, Jianmin Wang, and Philip~S. Yu.
\newblock Spatiotemporal pyramid network for video action recognition.
\newblock In {\em CVPR}, 2017.

\bibitem{wu2016multi}
Z.~Wu, Y.-G. Jiang, X.~Wang, H.~Ye, and X.~Xue.
\newblock Multi-stream multi-class fusion of deep networks for video
  classification.
\newblock In {\em ACM Multimedia}, 2016.

\bibitem{human_still_latent}
W.~Yang, Y.~Wang, and G.: Mori.
\newblock Recognizing human actions from still images with latent poses.
\newblock In {\em CVPR}, 2010.

\bibitem{graph_matching}
B.~Yao and L~Fei-Fei.
\newblock Action recognition with exemplar based 2.5d graph matching.
\newblock In {\em ECCV}, 2012.

\bibitem{late_fusion_2012}
G.~Ye, D.~Liu, I.-H. Jhuo, and S.-F. Chang.
\newblock Robust late fusion with rank minimization.
\newblock In {\em CVPR}, 2012.

\bibitem{icmr15:eval2stream}
H.~Ye, Z.~Wu, R.-W. Zhao, X.~Wang, Y.-G. Jiang, and X.~Xue.
\newblock Evaluating two-stream cnn for video classification.
\newblock In {\em ACM ICMR}, 2015.

\bibitem{yunansigir2017}
Yunan Ye, Zhou Zhao, Yimeng Li, Long Chen, Jun Xiao, and Yueting Zhuang.
\newblock Video question answering via attribute-augmented attention network
  learning.
\newblock In {\em SIGIR}, 2017.

\bibitem{ZhangWWQW16}
Bowen Zhang, Limin Wang, Zhe Wang, Yu~Qiao, and Hanli Wang.
\newblock Real-time action recognition with enhanced motion vector cnns.
\newblock In {\em CVPR}, 2016.

\bibitem{zhu2016key}
Wangjiang Zhu, Jie Hu, Gang Sun, Xudong Cao, and Yu~Qiao.
\newblock A key volume mining deep framework for action recognition.
\newblock In {\em CVPR}, 2016.

\end{thebibliography}

\end{document}